\documentstyle[a4,12pt]{article}
\def\lsim{\raise0.3ex\hbox{$\;<$\kern-0.75em\raise-1.1ex\hbox{$\sim\;$}}}
\def\gsim{\raise0.3ex\hbox{$\;>$\kern-0.75em\raise-1.1ex\hbox{$\sim\;$}}}
\def\Frac#1#2{\frac{\displaystyle{#1}}{\displaystyle{#2}}}

\begin{document}
\begin{flushright}
KUNS-1696\\
FTUV-00-11-15, IFIC-00-71\\
\end{flushright}

\begin{center}
{\Large \bf Restricted flavor structure of soft SUSY breaking trilinear 
couplings}
\vskip 0.75cm
{Tatsuo Kobayashi$^1$ and Oscar Vives$^2$}\\
\vspace*{2mm}\small{\textit{$^1$ Department of Physics, Kyoto University,
Kyoto 606-8502, Japan}}\\
\vspace*{2mm}\small{\textit{$^2$Departament de F\'{\i}sica Te\`orica and IFIC,
Universitat de Val\`encia-CSIC}\\
\textit{E-46100, Burjassot (Val\`encia), Spain}}\\
\vspace*{1.5cm}

\begin{abstract}
{We analyze the flavor structure of the trilinear couplings in the different
theoretical models
of SUSY breaking. We generically obtain $A_{ij}= A^L_i + A^R_j$ in all the 
models examined. In fact, this is the rigorous form when SUSY breaking effects
appear through the K\"ahler metric of chiral fields or through wave--function 
renormalization. Indeed, the low--energy phenomenological requirements from 
the absence of charge and color breaking minima and the measurements in flavor 
changing neutral current (FCNC) observables strongly favor this restricted 
form of the trilinear matrices. As a straightforward consequence the 
number of unknown parameters associated with the trilinear couplings is 
decreased in a factor of 2.}
\end{abstract}

\end{center}

\thispagestyle{empty}

\newpage
\section{Introduction}

The large popularity of Supersymmetry (SUSY) as an ingredient of many 
extensions of the Standard Model (SM) is due both to its ability to deal with 
several theoretical problems that the SM faces when it is stretched up to 
high energies
and to the fact that it naturally appears in most of the theories that
attempt to include gravity at the quantum level.
For these reasons, it is generally believed that global Supersymmetry must be 
discovered in the neighborhood of the electroweak scale with the new hadronic 
colliders.

However, from the phenomenological point of view, even the Minimal 
Supersymmetric extension of the SM (MSSM), the simplest supersymmetrization 
of the SM with no additional particle content, contains a host of free 
parameters related to the unknown SUSY soft--breaking terms. 
These soft--breaking terms, or more exactly, the trilinear couplings 
$Y^{A a}_{ij}$ and scalar mass matrices $m_{ij}$, have additional flavor 
structures besides the usual Yukawa matrices $Y^a_{ij}$. 
In fact, the low--energy phenomenology is strongly dependent on these new 
flavor structures. In the presence of strict flavor--universality, only 
observables with a dominant chirality changing contribution (i.e. electric 
dipole moments (EDM), $b \rightarrow s \gamma$ $\dots$) are sensitive to new 
supersymmetric contributions \cite{flavor,univ}.
Still, it has been recently shown that non--universality of the $A$--matrices 
is very relevant in the low energy observables \cite{masiero,n-uni-A,EDMfree,
KvsB} and it can raise a large contribution to $\varepsilon^\prime/\varepsilon$
without conflicting with EDM constraints \cite{EDMfree}. From this point of
view, the effects of generic non--universal soft--terms on indirect 
searches both in the kaon and the B systems was considered in Ref.~\cite{KvsB}.
 
On the other hand, regarding model building, several SUSY breaking and 
mediation mechanisms, e.g. supergravity--mediation \cite{SUGRA} which includes 
string--inspired supergravity theories\footnote{ For example, see Ref. 
\cite{string-SG} for heterotic models and Ref. \cite{typeI} for type I 
models.}, gauge--mediation \cite{GR} and anomaly--mediation 
\cite{randall,giudice,pomarol}\footnote{The application of pure anomaly 
mediation to the supersymmetric standard model has the tachyonic slepton 
mass problem. This problem can be solved by adding a universal soft 
scalar mass \cite{randall}. Another solution is $D$--term contributions 
to scalar masses which is quite interesting 
because it does not change the renormalization group flow \cite{AM-D}, 
one of the characteristic features of anomaly mediation.},  
have been proposed and each mechanism predicts characteristic flavor 
structures and sparticle spectrum. In this paper, we take advantage of this
complementary information to improve our knowledge of the soft--breaking terms,
and therefore of the complete MSSM. 
In particular we concentrate on the flavor structure of the $A$--terms.
In a purely phenomenological counting, the quark sector has in total 
$2\times 3\times 3 (=18)$ complex free parameters for the $A$--matrices, 
$A^u_{ij}$ and $A^d_{ij}$ associated with the up and down sector Yukawa 
matrices. However, we show that in quite generic 
models of SUSY breaking $A_{ij} = A^L_i + A^R_j$. That is, the quark
sector has only $3+3+3(=9)$ complex parameters for 
$A^u_{ij}$ and $A^d_{ij}$. In fact, this structure of the $A$--matrices 
guarantees the proportionality of the different trilinear couplings to the 
mass of one of the fermionic partners of the squarks involved. 
Furthermore, the magnitudes of the $A$--parameters are order of the gaugino 
and soft scalar masses. Due to these features, many of the phenomenological 
bounds on the so--called $LR$ Mass Insertions (MI) \cite{MI} are naturally 
avoided, as well as charge and color breaking minima and directions 
unbounded from bellow \cite{casas}.

In section 2 we analyze the structure of the $A$--matrices in several models of
SUSY breaking. We then discuss its phenomenological implications in section 3 
through discussions on Yukawa matrices and trilinear couplings.
Finally, in section 4 we present our conclusions.

\section{Generic form of $A$-matrices}

In this section, we analyze the form of the $A$-matrices in the main models of
SUSY breaking, i.e. supergravity mediation, gauge mediation and anomaly 
mediation. We show that the generic form of these matrices is, 
\begin{equation}
A_{ij} = A^L_i + A^R_j,
\label{A-1}
\end{equation}
which implies at the SUSY breaking scale,
\begin{equation}
Y^A_{ij} = Y_{ij}A_{ij} = 
\left(\begin{array}{ccc}
 &  &  \\  & Y_{ij} &  \\  &  & \end{array}
\right) \cdot 
\left(\begin{array}{ccc}
A^R_{1} & 0 & 0 \\ 0 & A^R_{2} & 0 \\ 0 & 0 & A^R_{3} \end{array}
\right) + 
\left(\begin{array}{ccc}
A^L_{1} & 0 & 0 \\ 0 & A^L_{2} & 0 \\ 0 & 0 & A^L_{3} \end{array}
\right) \cdot 
\left(\begin{array}{ccc}
 &  &  \\  & Y_{ij} &  \\  &  & \end{array}
\right) ,
\label{A-2}
\end{equation}
in matrix notation. This corresponds to a reduction in a factor of 2 of the 
number of free parameters in the trilinear couplings in the quark sector:
from 18 to 9 complex parameters (something similar happens in the leptonic 
sector).
We discuss the supergravity--mediation in section 2.1.
In section 2.2 we analyze models where SUSY breaking effects are induced 
through the spurion formalism. This mechanism includes both 
gauge--mediation and anomaly--mediation.

\subsection{Supergravity mediation}
We consider a supergravity model including, on one hand, chiral matter 
fields, $\Phi_{L}^i$, $\Phi_{R}^i$ and the Higgs field $H$, 
which remain light after SUSY breaking and we denote altogether as 
$\Phi^i$, and on the 
other hand some moduli fields $\Phi^a$, which have large vacuum expectation 
values breaking SUSY and then decouple from the low--energy physics.
The supergravity Lagrangian is written in terms of the
K\"ahler potential, the superpotential and the gauge kinetic 
functions. The K\"ahler potential provides kinetic terms of chiral fields 
while the superpotential includes both the Yukawa couplings and the $\mu$-term 
as well as a nonperturbative superpotential leading SUSY breaking.
We start with the following K\"ahler potential $K$ and 
the superpotential $W$,
\begin{eqnarray}
K &=& \tilde K(\Phi^a,\bar \Phi_a)+ K^i_i(\Phi^a,\bar \Phi_a)
|\Phi^i|^2, \\
W &=& \tilde W(\Phi^a)+ \hat Y_{ij}\Phi_{L}^i\Phi_{R}^j H,
\end{eqnarray}
with $\tilde W(\Phi^a)$ the nonperturbative superpotential 
leading to SUSY breaking.
The scalar potential is obtained as 
\begin{equation}
V = F^I F_J K_I^J - 3 e^G,
\end{equation}
where $\Phi^I$ denotes any of the fields $\Phi^i$ and $\Phi^a$, 
$F^I$ are the $F$-terms of the field $\Phi^I$ and $G=K+\ln|W|^2$.
The fields $\Phi^a$ develop their vacuum expectation values and their 
$F$-terms contribute to SUSY breaking. Then, taking the flat limit, 
we can expand the scalar potential around non--vanishing 
$\Phi^a$ and $F^a$,
\begin{equation}
V=V_0 + \left|{\partial W_{eff} \over \partial \phi_i}\right|^2 + 
{1 \over 2}m^2_i|\phi^i|^2K^i_i+( Y^A_{ij}\phi^i_L \phi^j_R h 
\prod_{\ell}(K^\ell_\ell)^{1/2} + h.c.) + \cdots,
\label{s-V}
\end{equation}
where $\phi^i$$(h)$ are the scalar component of $\Phi^i$$(H)$, and 
$\prod_{\ell} K^\ell_\ell$ denotes the product of 
normalization constants of the fields, $\Phi^i_L$,  $\Phi^i_R$ and $H$. 
The first term in the RHS of Eq.~(\ref{s-V}) is the vacuum energy obtained as 
$V_0= F_a F^b K^a_b - 3 m_{3/2}^2$ with the gravitino mass 
$m_{3/2}^2 \equiv e^G$.
The second term corresponds to globally supersymmetric contributions to the
scalar potential. Notice that we use $W_{eff}$, the effective superpotential 
in the global SUSY basis in terms of the canonically normalized fields. In this
basis the Yukawa couplings $Y_{ij}$ are obtained as,
\begin{equation}
|Y_{ij}|^2 = e^{\tilde K}|\hat Y_{ij}|^2 \prod_{\ell} (K^\ell_\ell)^{-1}.
\end{equation}
The last two terms in Eq.~(\ref{s-V}) are SUSY breaking terms, that is, 
the soft scalar mass $m_i^2$ and the trilinear scalar coupling $Y^A_{ij}$ 
which are obtained \cite{soni,KL}
\begin{eqnarray}
m_i^2 &=& m^2_{3/2} - F^a F_b\partial_a \partial^b(\ln K^i_i)+V_0,
\label{soft-m}\\
Y^A_{ij} &=& Y_{ij}F^a\partial_a(\tilde K - 
\ln \prod_{\ell} (K^\ell_\ell) ) +\Delta_{ij},
\label{soft-h}
\end{eqnarray}
where $\partial_a$ denotes $\partial /\partial \Phi^a$. In this expression we
obtain diagonal soft scalar masses. This results from the fact that we started 
with diagonal K\"ahler metric.
Generic supergravity theories can lead to non--vanishing off--diagonal 
K\"ahler metric $K^i_j$ $(i \neq j)$ for different flavors.
However, such mixing does not appear in superstring models at leading 
order. These models have additional continuous and discrete 
symmetries other than the SM gauge symmetry, and 
under these symmetries each massless state has definite charges.
Different flavors are distinguished by these extra charges 
and so, off--diagonal metric is forbidden.
When these extra symmetry are broken, non--vanishing off--diagonal K\"ahler
metric can appear. In the following, we assume that such off--diagonal 
K\"ahler metric induced by symmetry breaking is suppressed enough 
\cite{string-SG}.

In Eq.(\ref{soft-h}) we have included a term $\Delta_{ij}$ which denotes 
the term proportional to derivatives of $\hat Y^{ij}$ by the fields $\Phi_a$, 
$F_a \partial \hat Y^{ij}/ \partial \Phi_a$. Although these terms are perfectly
possible from the point of view of supergravity, indeed a large class of 
string--inspired supergravity theories satisfy the 
condition of field-independent Yukawa couplings. For instance, Yukawa 
couplings $\hat Y_{ij}$ are constants in heterotic Calabi--Yau models in the 
large $T$ limit and type--I models \cite{typeI}.
In these theories, the Yukawa couplings are not functions of the fields 
$\Phi_a$ with non--vanishing F--terms $F_a \neq 0$, and so the $(i,j)$ entry 
of the $A$-matrix is always written in the from 
\begin{equation}
A_{ij} = A^L_i + A^R_j +A_0,
\end{equation}
where 
\begin{eqnarray}
A^L_i &=& - F_a \partial^a \ln K^i_i, \\
A^R_j &=& - F_a \partial^a \ln K^j_j, \\
A_0 &=& F_a \partial^a (\tilde K -\ln K^H_H).
\end{eqnarray}
If we redefine $A^L_i$ and  $A^R_j$ absorbing $A_0$, then the resultant form 
exactly corresponds to eqs.(\ref{A-1}), (\ref{A-2}).
For example, we redefine $A'^R_j = A^R_j+A_0$ and then 
the generic form of $A$-matrix is 
\begin{equation}
A_{ij} =  A^L_i + A'^R_j ,
\end{equation}
under the condition that the Yukawa couplings do not depend 
on the fields $\Phi_a$ with 
non--vanishing F--terms $F_a \neq 0$.

More generally, in some models the Yukawa couplings can be field--dependent
(although it is often difficult to calculate field--dependent parts of 
Yukawa couplings and it is believed that such parts are suppressed compared 
with the constant term). One of the examples where an explicit calculation 
can be done is heterotic orbifold models. There are two types of 
closed string sectors, untwisted and twisted strings.
In addition, each twisted sector is assigned to 
a fixed point $f_i$ on the orbifold.
Yukawa couplings of untwisted sectors are 
constants, and Yukawa couplings of twisted sectors 
associated to the same fixed point are also constants.
Thus, these two couplings do not lead to non--vanishing $\Delta_{ij}$.
Yukawa couplings of twisted sectors with different fixed points are 
obtained by the world-sheet instanton action and they 
depend on the moduli field T, $Y_{ij} \sim e^{-a_{ij}T}$, 
where a vacuum expectation value of $T$ 
corresponds to the compactification size \cite{HV,DFMS}.
Combinations of fixed points for allowed Yukawa couplings are 
unique, and so the coupled fixed point is determined 
uniquely with the other two points chosen.
Thus we can write the coefficient 
$a_{ij} \propto f_{Li}- f_{Rj}$ \cite{DFMS}, where $f_{Li}$ ($f_{Rj}$) 
is the fixed point associated to the field $\Phi^i_{L}$ ($\Phi^i_{R}$).
So, in this case, the contribution due to the field--dependent 
Yukawa coupling $\Delta_{ij}$ also takes the form 
$\Delta_{ij}= \Delta^{L}_i + \Delta^R_j$.
Note that even if the Yukawa couplings depend on the moduli fields 
the corresponding $A$-matrix has still the form in Eqs.~(\ref{A-1},\ref{A-2}) 
in two cases: $F$-terms of $T$ are not dominant or the dependence of the 
Yukawa couplings can be factorized as in the case of heterotic orbifold models.

Another example of field--dependent Yukawa couplings is 
given by Calabi--Yau models without the large $T$ limit.
We consider the case with one moduli field $T$.
Here Yukawa couplings are obtained as $Y_{ij} \sim C + d e^{- a T}$, 
where $C$, $d$, $a$ are constants, and Yukawa couplings are universal.
Hence, this situation simply provides a flavor--universal
correction $\Delta_{ij}$.
With only these couplings of ${\cal{O}}(1)$ one can not lead to 
realistically hierarchical Yukawa matrices, either.
We need higher dimensional operators like the Froggatt--Nielsen 
mechanism \cite{FN} to derive realistic Yukawa matrices, which we discuss
in the next section. However, coupling strengths of higher dimensional 
operators have not been calculated completely for Calabi--Yau models 
or orbifold models and possible corrections given by these
operators would be suppressed, in any case, although it certainly deserves
further analysis.

\subsection{Gauge--mediation and anomaly--mediation: 
spurion formalism}

We consider a global SUSY model with chiral matter fields, 
$\Phi^i_{L}$ and $\Phi^i_{R}$ and the Higgs field $H$.
We start with the following renormalized Lagrangian at the scale $\mu$,
\begin{eqnarray}
{\cal L} &=&   \int d^4 \theta\ 
Z(\Lambda_M,\mu)_i\  \Phi_i^\dag e^V \Phi^i \nonumber \\
         &+& \int d^2\theta\ S(\Lambda_M,\mu)_\alpha\ W^\alpha W_\alpha 
+   \int d^2 \theta\ W(\Phi) + {\rm h.c.}  ,\nonumber\\
W(\Phi) &=& Y_{ij} \Phi_{L}^i \Phi_{R}^j H, 
\end{eqnarray}
where $Z(\Lambda_M,\mu)_i$ and $S(\Lambda_M,\mu)_\alpha$ are the 
wave--function 
renormalization constant of $\Phi^i$ and the renormalized coupling for the 
gauge multiplet $V$ respectively. Moreover $W^\alpha$ is the field strength 
superfield of $V$. In this equation, $\Lambda_M$ is simply a threshold 
energy scale, where some matter fields become heavy and decouple, and 
the $\beta$-function changes. Alternatively we can consider $\Lambda_M$ 
as a non--dynamical field determined by a vacuum expectation value of a 
superfield $M$. Then, the superfield $M$ acquires a vacuum expectation 
value along the scalar and auxiliary components,
\begin{equation}
<M> = \Lambda_M + \theta^2 F_M,
\end{equation}
and so, SUSY is broken by F--term of $M$.
We assume that the effective Lagrangian is still valid 
after replacing $\Lambda_M^2 \rightarrow |M|^2$ 
in $Z(\Lambda_M,\mu)_i$ and $S(\Lambda_M,\mu)_\alpha$ even 
with non--vanishing $F_M$.
Thus, SUSY breaking effects due to $M$ appear through the
$\Lambda_M$--dependence of the wave--function renormalization $Z$ 
as well as $S$. One can consider the case with more than one threshold 
in a similar way. This is the mechanism used in gauge--mediated SUSY 
breaking \cite{GR}. Hence, the expression for the soft SUSY breaking A--term
is,
\begin{eqnarray}
A(\mu)_{ij} &=& \sum_{\ell } 
{\partial ln Z_\ell(M,\bar M,\mu) \over 
\partial ln M} {F_M \over \Lambda_M}, 
\end{eqnarray}
where the summation is taken for the wave--functions of 
$\Phi_L^i$, $\Phi_R^j$ and $H$.
Thus, within this framework, 
the $A$--matrix $A_{ij}$ is the summation of the three parts, i.e. 
$\Phi^L_{i}$--dependent part, $\Phi^R_{j}$--dependent part and 
the Higgs part. The Higgs part is universal for any family.
So, it is straight--forward to obtain the form in Eqs.~(\ref{A-1},\ref{A-2}) 
after we absorb the Higgs part into the $\Phi^R_{j}$--dependent part, for 
instance.

In the case of anomaly--mediation, we consider $\Lambda_M$ as the cut--off 
scale of the MSSM, above which 
our 4-dimensional field theory is not valid, that is, 
the cut-off $\Lambda_M$ could be the string scale, compactification
scale or the breaking scale of conformal symmetry.
All the expressions above remain 
exactly the same. Thus, also in anomaly--mediation, 
we have the same form of Eqs.~(\ref{A-1},\ref{A-2}).
Furthermore, we always have this form of the $A$--matrices in a generic case 
where SUSY breaking effects appear only in wave--function renormalization 
through loop effects.\footnote{See for example Ref. \cite{arkani-hamed}. 
As a further example, this class of SUSY breaking mediation 
mechanisms also includes the case where SUSY breaking appears through loop 
effects due to Kaluza--Klein modes in extra dimensional models \cite{KK-med}.
Even in such case we have the form of $A$--matrices in 
Eqs.~(\ref{A-1},\ref{A-2}).}
Moreover, natural magnitudes of $A_{Lj}$ and $A_{Rj}$ are 
obtained as $A_{Lj}, A_{Rj} \leq {\cal{O}}(M_\alpha)$, with $M_\alpha$ the 
gaugino mass.

\section{Yukawa textures and phenomenological implications}

In the previous section, we have seen that 
the form of A--matrices, (\ref{A-1}), (\ref{A-2}) 
is obtained in quite generic case.
If we apply this form to the MSSM, the $A^u_{ij}$ and $A^d_{ij}$ matrices are 
written as,
\begin{equation}
A^u_{ij} = A^Q_i + A^U_j, \qquad A^d_{ij} = A^Q_i + A^D_j.
\label{A-ud}
\end{equation} 
This structure has very important phenomenological effects.  
In first place, it implies an important reduction on the number of free 
parameters associated to the trilinear coupling matrices. 
We fix the eighteen complex 
matrix elements in the quark sector with only nine complex parameters.
Secondly, the low energy flavor changing (FCNC) phenomenology sets very 
stringent bounds on generic $LR$ MI \cite{MI} and moreover, the absence of 
charge and color breaking minima and directions unbounded from bellow 
\cite{casas}
constrains strongly the $LR$ sfermion mixing. All these phenomenological  
requirements imply that the structure of the trilinear couplings, $Y^A_{ij}$,
goes beyond the usually assumed form $Y^A_{ij}=A_{ij} Y_{ij}$. Indeed, we
show bellow that the trilinear couplings in Eqs.~(\ref{A-1},\ref{A-2}) obtained
in generic SUSY breaking models naturally fit in the low energy scenario 
as required by the available phenomenological constraints.

To do this, we first make some general remarks on the Yukawa matrices in these
models because these Yukawa matrices are the additional ingredient in the 
flavor structure of the whole trilinear couplings, $Y^A_{ij}$. 
In a model independent way, we can write the Yukawa 
matrices in the basis of diagonal sfermion masses as follows,
\begin{equation}
v_1\, Y_d = {K^{U_L}}^\dagger\cdot V_{CKM}\cdot  M_d\cdot K^{D_R}, \qquad 
v_2\, Y_u = {K^{U_L}}^\dagger\cdot M_u\cdot K^{U_R},
\end{equation}
where $v_1$ ($v_2$) is the vacuum expectation value of the 
down (up) sector Higgs field, 
$M_d$ and $M_u$ are diagonal quark mass matrices, $V_{CKM}$ the 
Cabibbo--Kobayashi--Maskawa (CKM) mixing matrix and $K^{U_L}$, $K^{D_R}$, 
$K^{U_R}$ general $3 \times 3$ unitary matrices. 
These matrices measure the flavor misalignment among, $u_L$--$\tilde{Q}_L$, 
$d_R$--$\tilde{d}_R$ and $u_R$--$\tilde{u}_R$ respectively and their 
structure depends on the particular theory of supersymmetry breaking 
and Yukawa flavor that we consider. Hence, we can not give the final 
texture of these matrices, but we can still discuss several generic 
aspects of their flavor structures. Gauge interactions are completely 
flavor--blind and the existence of three different generations must be
understood in terms of additional symmetries. In underlying theory, 
as for instance superstring inspired models,
all allowed Yukawa couplings (at the string scale) are naturally order 1.
Other couplings can only be obtained through higher dimensional operators and 
hence are suppressed hierarchically through a mechanism analogous to the
well--known Froggatt--Nielsen mechanism \cite{FN}.
That is, within this mechanism, some extra symmetries, e.g. $U(1)$ symmetries 
and/or discrete symmetries, are assumed and 
extra charges are assigned to the MSSM fields and extra Higgs fields 
$\chi_k$. Then, after $\chi_k$ develop their vacuum expectation values, 
these higher dimensional operators generate effective Yukawa couplings, 
$y_{ij}({\chi_k / M})^{n^u_{ij}}Q_i U_j H_2$ and 
$y_{ij}({\chi_k / M})^{n^d_{ij}}Q_i D_j H_1$ .
In a similar way, stringy selection rules of Yukawa couplings and higher 
dimensional operators\footnote{See e.g. Ref. \cite{st-rule} 
for stringy selection rules of higher dimensional operators in heterotic 
orbifold models.} can be understood in terms of discrete symmetries. In fact,
these symmetries determine completely the flavor structure both in the
Yukawa and soft--breaking sectors within the framework of 
string--inspired supergravity.

To reproduce correctly the observed hierarchy of masses and mixings, the 
$Y^u_{3 3}$ must be given as a 3--point coupling with $n^u_{33}=0$, 
because experimentally the top Yukawa coupling is of ${\cal{O}}(1)$. All other 
couplings are suppressed by 
$\epsilon^{u,d}_{i,j}=({\chi_k / M})^{n^{u,d}_{ij}}$
depending on the charges $n^{u,d}_{ij}$ with $({\chi_k / M})<< 1$.
The up Yukawa matrix would be then, 
\begin{equation}
Y^{u} \simeq Y_t \left(\begin{array}{ccc}
\epsilon^u_{1,1}& \epsilon^u_{1,2}  &  \epsilon^u_{1,3} \\
\epsilon^u_{2,1}& \epsilon^u_{2,2}  &  \epsilon^u_{2,3} \\
\epsilon^u_{3,1}  &\epsilon^u_{3,2}  & 1 \end{array}
\right). 
\label{Y-u}
\end{equation}

At this point, it is interesting to comment the possibility of obtaining a
maximal mixing \cite{everett}. 
For instance, in the left--handed sector, the maximal $(3,i)$ mixing could be 
realized for $\epsilon^u_{i,3} = 1$.
The condition $n^u_{33}=0$ fixes uniquely the extra charges of $Q_3$ 
once the charges of $U_3$ and $H_2$ are chosen definitely. 
Thus, the condition $\epsilon^u_{i,3} = 1$ i.e., $n^u_{i3}=0$, requires 
that both $Q_1$ and $Q_2$ should have the same extra charges as $Q_3$.
Hence, nothing can distinguish the three $Q_i$.
Under such situation, it is clear that both 
$A^Q_i$ and soft scalar masses of $Q_i$ are universal and there is no new 
flavor structure in the left--sector. We can rotate freely this democratic 
couplings to a single entry in the $(3,3)$ element and we have Eq.~(\ref{Y-u})
with $\epsilon^{u}_{i,j}<< 1$. So, in this final basis 
we have again a hierarchical structure and the observable mixing angles will
be small.

Once we assume a hierarchical structure, there is a simple
relation among Yukawa elements and mixing angles at first order:
$K^{{U_L}}_{3 2} = {\epsilon^u_{2,3}}^*$,  
$K^{{U_L}}_{3 1} = {\epsilon^u_{1,3}}^*$, 
$K^{{U_R}}_{3 2} = \epsilon^u_{3,2}$ and $K^{{U_R}}_{3 1} = \epsilon^u_{3,1}$. 
This suppression is then transmitted to the $K^{U_{L,R}}_{i 3}$ and $K^{D_L}$ 
by unitarity of the mixing matrices and by the CKM mixing respectively.
In principle, the $K^{D_R}$ mixings are not strongly constrained except in
the large $\tan \beta$ regime, where only $Y^d_{3 3} \simeq 1$ and all other 
elements must be suppressed. In summary, we have discussed the form (\ref{Y-u}) 
with suppressed factors $\epsilon^u_{i,j}$ and found 
$K^{(U,D)_L}_{3i},K^{(U,D)_L}_{i3},K^{U_R}_{3i},K^{U_R}_{i3} << 1$.
However, explicit values of $\epsilon^u_{i,j}$ are model--dependent 
and it is difficult to estimate them for generic case.
Henceforth we use the following approach in our estimates: we do not 
expect an accidental cancellation to obtain the CKM matrix, 
$V_{CKM}=K^{U_L}{K^{D_L}}^\dagger$.
So, this means for instance, $\mbox{Max}(K^{U_L}_{13},K^{D_L}_{13}) 
\sim \lambda^3$, with $\lambda$ the Cabibbo angle.
Notice that this assumption implies a further step which could be
easily circumvented in some models, but we consider it a natural
feature in most of the models that we analyze here. 
Still, we comment our results with and without this assumption.

The requirement of the absence of charge and color breaking minima and 
directions unbounded from bellow sets strong constraints on 
off--diagonal elements of the $Y^A_{ij}$ matrix. In fact from charge and 
color breaking minima the following bounds are required \cite{casas},
\begin{eqnarray}
\label{CCB}
&|Y^{A u}_{ij}|^2 \leq Y_{u_k}^2 \Big(m_{U_{L_i}}^2 + m_{U_{R_j}}^2 + 
m_2^2\Big), & \\
&|Y^{A d}_{ij}|^2 \leq Y_{d_k}^2 \Big(m_{D_{L_i}}^2 + m_{D_{R_j}}^2 + 
m_1^2\Big), &  k = \mbox{Max} (i,j), \nonumber
\end{eqnarray}
and similarly from directions unbounded from bellow,  
\begin{eqnarray}
\label{UFB}
&|Y^{A u}_{ij}|^2 \leq Y_{u_k}^2 \Big(m_{U_{L_i}}^2 + m_{U_{R_j}}^2 + 
m_{E_{L_p}}^2+ m_{E_{R_q}}^2\Big), & k = \mbox{Max} ( i, j),\ p \neq q \\
&|Y^{A d}_{ij}|^2 \leq Y_{d_k}^2 \Big(m_{D_{L_i}}^2 + m_{D_{R_j}}^2 + 
m_{\nu_m} ^2\Big), &  k = \mbox{Max} (i,j),\ m \neq i,j \nonumber
\end{eqnarray} 
in the basis where Yukawa couplings, $Y_{u_k}$, $Y_{d_k}$, are diagonal. It 
is important to notice
that these bounds are indeed competitive and in many cases more stringent than 
the corresponding FCNC bounds \cite{MI,casas}. However, from 
Eqs.~(\ref{CCB},\ref{UFB}) it is evident that for $A_{ij}$ elements of the 
same order of the scalar masses, the main condition these bounds require is 
precisely that the masses of the fermionic partners of the squarks involved
set the scale of the coupling, which we naturally find in the SUSY breaking 
models that we analyze.       
Note that, as can be seen bellow, the LHS's in Eqs.~(\ref{CCB}) and 
(\ref{UFB}) usually include additional suppression factors due to the 
diagonalizing $K$--matrices in the Yukawa--diagonal basis.

Similarly FCNC processes set very stringent bounds on generic $LR$ squark 
mixing matrices \cite{MI}. Nevertheless, when we take into account the
proportionality to the fermion masses these constraints are largely relieved.  
Following references \cite{EDMfree,KvsB}, and using Eq.~(\ref{A-ud}), an 
estimate\footnote{The main RGE effects are either flavor--universal or 
flavor--diagonal in the basis of diagonal Yukawas \cite{RGE}.} of the off--diagonal $LR$ MI at $M_W$ can be obtained as,  
\begin{eqnarray}
\label{LRMI}
(\delta^{d}_{LR})_{i\neq j}&\ =\ \Frac{1}{m^{2}_{\tilde{q}}}\ \Big( m_j\ 
(A^Q_2 -  A^Q_1)\ K^{D_L}_{i 2} {K^{D_L}_{j 2}}^*\ +\  m_j\ (A^Q_3 -  
A^Q_1) K^{D_L}_{i 3} {K^{D_L}_{j 3}}^* \nonumber \\
&\ \ +\  m_i\  (A^D_2 -  A^D_1)\ K^{D_R}_{i 2} {K^{D_R}_{j 2}}^*\ +\  m_i\ 
(A^D_3 -  A^D_1)\ K^{D_R}_{i 3} {K^{D_R}_{j 3}}^* \Big) ,
\end{eqnarray}
where $m_{\tilde{q}}$ is the average squark mass.
The value $(\delta^{d}_{LR})_{i j}$ depends on  
$K^{D_L}$ and $K^{D_R}$ and hereafter we neglect small masses as $m_d/m_s$ or
$m_s/m_b$.

If we analyze the MI which contribute in the Kaon system, we obtain
\begin{equation}
\label{LRKaon1}
(\delta^{d}_{LR})_{1 2}\ \simeq\ \Frac{m_s}{m_{\tilde{q}}}\ 
\Frac{(A^Q_2 -  A^Q_1)}{m_{\tilde{q}}}\ K^{D_L}_{1 2} {K^{D_L}_{2 2}}^*
\end{equation} 
and, 
\begin{equation}
\label{LRKaon2}
(\delta^{d}_{LR})_{2 1}\ \simeq\ \Frac{m_s}{m_{\tilde{q}}}\ 
\Frac{(A^D_2 -  A^D_1)}{m_{\tilde{q}}}\ K^{D_R}_{1 2} {K^{D_R}_{2 2}}^*.
\end{equation} 
In this expression we can see clearly what is important to obtain a large
MI. Both, non--universality of $A$--matrices and
a sizeable mixing among squark generations are needed. 
In this case we have $K^{D_L}_{1 2} = {\cal{O}}(\lambda)$ and 
$(A^Q_2 -  A^Q_1)/m_{\tilde{q}} = {\cal{O}}(1)$, and this is enough to give 
rise to a very 
sizable contribution to $\varepsilon^\prime/\varepsilon$ \cite{EDMfree,KvsB}. 
As explained above, the right handed mixings are in principle unconstrained 
but, in any case, it is very difficult to expect a larger mixing in the
$1$--$2$ sector given that, due to unitarity, the maximal
value for $K^{D_R}_{1 2} {K^{D_R}_{2 2}}^* = 0.5$ (to be compared with
$K^{D_L}_{1 2} {K^{D_L}_{2 2}}^* = 0.22$). Hence, can be 
$(\delta^{d}_{LR})_{2 1}$ at most a factor 2 larger than the   
$(\delta^{d}_{LR})_{1 2}$. In summary, thanks to the high sensitivity of 
$\varepsilon^\prime/\varepsilon$ and to the presence 
of a large mixing among the first two generations, these MI can still have an 
observable effect, even overcoming the large mass suppression 
$m_s/m_{\tilde{q}}$ 

Similarly, in the neutral $B$ system, contributions to the $B_d$--$\bar{B}_d$ 
mixing parameter, $\Delta M_{B_d}$ are controlled by,
\begin{equation}
\label{LRbottom}
(\delta^{d}_{LR})_{1 3}\ \simeq\ \Frac{m_b}{m_{\tilde{q}}}\ 
\Frac{(A^Q_3 -  A^Q_1)}{m_{\tilde{q}}}\ K^{D_L}_{1 3} {K^{D_L}_{3 3}}^*,
\end{equation}  
and
\begin{equation}
\label{RLbottom}
(\delta^{d}_{LR})_{3 1}\ \simeq\ \Frac{m_b}{m_{\tilde{q}}}\ 
\Frac{(A^D_3 -  A^D_1)}{m_{\tilde{q}}}\ K^{D_R}_{1 3} {K^{D_R}_{3 3}}^*.
\end{equation} 
A natural value for the mixing angles is $K^{D_L}_{13} {K^{D_R}_{3 3}}^* 
\simeq \lambda^3$, which together with $(A^Q_3 -  A^Q_1)/m_{\tilde{q}} \sim 
{\cal{O}}(1)$ implies $(\delta^{d}_{LR})_{1 3}\simeq 5 \times
10^{-5}$, much smaller the MI $LR$ bounds for $\tilde{b}$--$\tilde{d}$
transitions. Even for $K^{D_R}_{1 3} {K^{D_R}_{3 3}}^* \lsim 0.5$ we obtain 
$(\delta^{d}_{LR})_{1 3}\simeq 3 \times 10^{-3}$, roughly
one order of magnitude too small to saturate $\Delta M_{B_d}$\footnote{Another 
interesting MI is $(\delta^{d}_{LR})_{3 2}$ which contributes to 
$b \rightarrow s \gamma$. However, in the case of hierarchical
Yukawa structures these MI are still small (see last paper in Ref. 
\cite{n-uni-A}).}.

As a result, we have shown that low energy phenomenology fits nicely
with the trilinear couplings in Eqs.~(\ref{A-1},\ref{A-2}). In fact this 
structure relieves the strong constraints from charge and color breaking
and most of the FCNC constraints. Still, as has been recently shown,
observables in the kaon sector are very sensitive to this trilinear couplings
\cite{masiero,n-uni-A,EDMfree}. 

\section{Conclusions}
In this work, we have studied the flavor structure in the soft SUSY breaking 
trilinear couplings. We have shown that we obtain $A_{ij}= A^L_i + A^R_j$ in a 
quite generic case, that is, when SUSY breaking effects appear through the 
K\"ahler metric of chiral fields or through wave--function renormalization due 
to loop effects. Furthermore, even in the known examples with the Yukawa 
couplings depending on the moduli fields with non--vanishing $F$--terms, as
for instance in heterotic orbifold models, this form is still maintained.
Then, we have investigated the phenomenological implications of this
form 
of the  
trilinear couplings. We have found that they naturally satisfy the conditions
required from the absence of charge and color breaking minima and directions 
unbounded from bellow. Similarly, we have found that they are safe
with 
most FCNC 
constraints with the only remarkable exception of 
$(\delta^{d}_{LR})_{12 (21)}$ from $\varepsilon^\prime/\varepsilon$.

\section*{Acknowledgments}
We thank the organizers of SUSY2K at CERN where this work was initiated. 
O.V. acknowledges financial support from the Marie Curie EC grants 
TMR-ERBFMBI CT98 3087 and MCFI-2000-0114 and partial 
support from spanish CICYT AEN-99/0692.


\begin{thebibliography}{99}

\bibitem{flavor}
D.~A.~Demir, A.~Masiero and O.~Vives,
Phys.\ Lett.\  {\bf B479}, 230 (2000) 
[hep-ph/9911337];
\\
D.A. Demir, A. Masiero and O. Vives, Phys.\ Rev.\ {\bf D61}, 075009 (2000)
[hep-ph/9909325].

\bibitem{univ}
P.~Cho, M.~Misiak and D.~Wyler,
Phys.\ Rev.\  {\bf D54}, 3329 (1996)
[hep-ph/9601360];
\\
S.~Baek and P.~Ko,
Phys.\ Lett.\  {\bf B462}, 95 (1999)
[hep-ph/9904283];
\\
A.~Ali and D.~London,
Eur.\ Phys.\ J.\  {\bf C9} (1999) 687
[hep-ph/9903535];
\\
A.~J.~Buras, P.~Gambino, M.~Gorbahn, S.~Jager and L.~Silvestrini,
hep-ph/0007313.

\bibitem{masiero}
A.~Masiero and H.~Murayama,
Phys.\ Rev.\ Lett.\  {\bf 83} (1999) 907
[hep-ph/9903363].

\bibitem{n-uni-A}
S.~A.~Abel and J.~M.~Frere,
Phys.\ Rev.\  {\bf D55} (1997) 1623
[hep-ph/9608251];
\\
S.~Khalil, T.~Kobayashi and A.~Masiero,
Phys.\ Rev.\  {\bf D60} (1999) 075003
[hep-ph/9903544];
\\
S.~Khalil and T.~Kobayashi,
Phys.\ Lett.\  {\bf B460} (1999) 341
[hep-ph/9906374];
\\
R.~Barbieri, R.~Contino and A.~Strumia,
Nucl.\ Phys.\  {\bf B578} (2000) 153
[hep-ph/9908255];
\\
K.~S.~Babu, B.~Dutta and R.~N.~Mohapatra,
Phys.\ Rev.\  {\bf D61} (2000) 091701
[hep-ph/9905464];
\\
M.~Brhlik, L.~Everett, G.~L.~Kane, S.~F.~King and O.~Lebedev,
Phys.\ Rev.\ Lett.\  {\bf 84} (2000) 3041
[hep-ph/9909480];
\\
D.~Bailin and S.~Khalil,
hep-ph/0010058.

\bibitem{EDMfree}
S.~Khalil, T.~Kobayashi and O.~Vives,
Nucl.\ Phys.\  {\bf B580} (2000) 275
[hep-ph/0003086].
\bibitem{KvsB}
A.~Masiero and O.~Vives,
hep-ph/0007320, to be published in Phys.\ Rev.\ Lett.\ 

\bibitem{SUGRA}
See for example,   
H.~P.~Nilles,
Phys.\ Rept.\  {\bf 110} (1984) 1.

\bibitem{string-SG}
A.~Brignole, L.~E.~Ib\'a\~nez and C.~Mu\~noz,
Nucl.\ Phys.\  {\bf B422} (1994) 125
[hep-ph/9308271];
\\
T.~Kobayashi, D.~Suematsu, K.~Yamada and Y.~Yamagishi,
Phys.\ Lett.\  {\bf B348} (1995) 402
[hep-ph/9408322];
\\
A.~Brignole, L.~E.~Ib\'a\~nez, C.~Mu\~noz and C.~Scheich,
Z.\ Phys.\  {\bf C74} (1997) 157
[hep-ph/9508258].

\bibitem{typeI}
L.~E.~Ib\'a\~nez, C.~Mu\~noz and S.~Rigolin,
Nucl.\ Phys.\  {\bf B553} (1999) 43
[hep-ph/9812397].

\bibitem{GR}
G.~F.~Giudice and R.~Rattazzi,
Phys.\ Rept.\  {\bf 322} (1999) 419
[hep-ph/9801271].

\bibitem{randall}
L.~Randall and R.~Sundrum,
Nucl.\ Phys.\  {\bf B557} (1999) 79
[hep-th/9810155].

\bibitem{giudice}
G.~F.~Giudice, M.~A.~Luty, H.~Murayama and R.~Rattazzi,
JHEP {\bf 9812} (1998) 027
[hep-ph/9810442].

\bibitem{pomarol}
A.~Pomarol and R.~Rattazzi,
JHEP {\bf 9905} (1999) 013
[hep-ph/9903448].

\bibitem{AM-D}
I.~Jack and D.~R.~Jones,
Phys.\ Lett.\  {\bf B482} (2000) 167
[hep-ph/0003081];
\\
M.~Carena, K.~Huitu and T.~Kobayashi,
hep-ph/0003187, to be published in Nucl. Phys. {\bf B}.


\bibitem{MI}
F.~Gabbiani, E.~Gabrielli, A.~Masiero and L.~Silvestrini,
Nucl.\ Phys.\ {\bf B477}, 321 (1996);
\\
J.~Hagelin, S.~Kelley and T.~Tanaka,
Nucl.\ Phys.\  {\bf B415}, 293 (1994).

\bibitem{casas}
J.~A.~Casas and S.~Dimopoulos,
Phys.\ Lett.\  {\bf B387}, 107 (1996)  
[hep-ph/9606237].


\bibitem{soni}
S.~K.~Soni and H.~A.~Weldon,
Phys.\ Lett.\  {\bf B126} (1983) 215.

\bibitem{KL}
V.~S.~Kaplunovsky and J.~Louis,
Phys.\ Lett.\  {\bf B306} (1993) 269
[hep-th/9303040].

\bibitem{HV}
S.~Hamidi and C.~Vafa,
Nucl.\ Phys.\  {\bf B279} (1987) 465.

\bibitem{DFMS}
L.~Dixon, D.~Friedan, E.~Martinec and S.~Shenker,
Nucl.\ Phys.\  {\bf B282} (1987) 13.

\bibitem{FN}
C.~D.~Froggatt and H.~B.~Nielsen,
Nucl.\ Phys.\  {\bf B147} (1979) 277.

\bibitem{arkani-hamed}
N.~Arkani-Hamed, G.~F.~Giudice, M.~A.~Luty and R.~Rattazzi,
Phys.\ Rev.\  {\bf D58} (1998) 115005
[hep-ph/9803290].

\bibitem{KK-med}
T.~Kobayashi and K.~Yoshioka,
hep-ph/0008069, to be published in Phys. Rev. Lett.

\bibitem{st-rule}
M.~Cveti\u{c},
Phys.\ Rev.\ Lett.\  {\bf 59} (1987) 1795;
Phys.\ Rev.\ Lett.\  {\bf 59} (1987) 2829;
\\
A.~Font, L.~E.~Ib\'a\~nez, H.~P.~Nilles and F.~Quevedo,
Nucl.\ Phys.\  {\bf B307} (1988) 109;
Phys.\ Lett.\  {\bf 210B} (1988) 101;
Phys.\ Lett.\  {\bf B213} (1988) 274;
\\
A.~Font, L.~E.~Ib\'a\~nez, F.~Quevedo and A.~Sierra,
Nucl.\ Phys.\  {\bf B331} (1990) 421;
\\
T.~Kobayashi,
Phys.\ Lett.\  {\bf B354} (1995) 264
[hep-ph/9504371];
Phys.\ Lett.\  {\bf B358} (1995) 253
[hep-ph/9507244].

\bibitem{everett}
L.~Everett, G.~L.~Kane and S.~F.~King,
JHEP {\bf 0008} (2000) 012
[hep-ph/0005204].

\bibitem{RGE}
N.K.~Falck, Z\ Phys.\ {\bf C30}, 247 (1986);
\\
S.~Bertolini, F.~Borzumati, A.~Masiero and G.~Ridolfi,
Nucl.\ Phys.\ {\bf B353}, 591 (1991).


\end{thebibliography}
\end{document}